STScI | SPACE TELESCOPE SCIENCE INSTITUTE

# JWST TECHNICAL REPORT

| Title: Characterization of the visit-to-visit Stability of the GR700XD Wavelength Calibration for NIRISS/SOSS Observations | Doc #: JWST-STScI-008571, SM-12<br>Date: 17 October 2023<br>Rev: - |
|---|---|
| Authors: Tyler Baines, Néstor Espinoza, Joseph Filippazzo, and Kevin Volk — Phone: (410) 338-4724 | Release Date: 10 November 2023 |

## 1. Abstract

When utilizing the NIRISS/SOSS mode on JWST, the pupil wheel (tasked with orienting the GR700XD grism into the optical path) does not consistently settle into its commanded position resulting in a minor misalignment with deviations of a few fractions of a degree. These small offsets subsequently introduce noticeable changes in the trace positions of the NIRISS SOSS spectral orders between visits. This inconsistency, in turn, can lead to variations of the wavelength solution. In this report, we present the visit-to-visit characterization of the NIRISS GR700XD Wavelength Calibration for spectral orders 1 and 2. Employing data from Calibration Program 1512 (PI: Espinoza), which intentionally and randomly sampled assorted pupil wheel positions during observations of the A-star BD+60-1753, as well as data from preceding commissioning and calibration activities to model this effect, we demonstrate that the wavelength solution can fluctuate in a predictable fashion between visits by up to a few pixels. We show that via two independent polynomial regression models for spectral orders 1 and 2, respectively, using the measured x-pixel positions of known Hydrogen absorption features in the A-star spectra and pupil wheel positions as regressors, we can accurately predict the wavelength solution for a particular visit with an RMS error within a few tenths of a pixel. We incorporate these models in PASTASOSS, a Python package for predicting the GR700XD spectral traces, which now allows to accurately predict spectral trace positions and their associated wavelengths for any NIRISS/SOSS observation.

## 2. Introduction

NIRISS/SOSS is a key JWST instrument mode for a wide range of science. In particular, its optical-to-near infrared wavelength range from 0.6 – 2.8 $\mu$m and moderate spectral resolution (R≈700 at 1.4 $\mu$m) makes it a prime mode for transiting exoplanet science (see Albert et al., 2023 and references therein). Such high-precision spectroscopic science, among other criteria, necessitates a comprehensive understanding of the instrument's stability, both intra- and inter-visit. While in-visit stability has already been studied at different levels through existing time-series observations (TSOs) presented by the scientific community (e.g., Fu et al., 2022; Feinstein et al., 2023; Radica et al., 2023; Holmberg & Madhusuhan, 2023; Madhusuhan et al. 2023; Lim et al., 2023), intra-visit stability investigations remain sparse. Such studies are vital for comprehending the limitations of the instrument during multiple-visit TSOs and evaluating the viability of science cases, such as long-term spectrophotometric monitoring of astrophysical sources.

**Table 1: List of the SOSS observation programs used to characterize the visit-to-visit wavelength solution including program PI, observation number, and their respective measured pupil wheel position angles given by PWCPOS keyword.**

| Program ID | Program PI | Observation Number | PWCPOS [deg] |
|---|---|---|---|
| 1091 | André Martel | 002 | 245.791 |
| 1536 | Karl Gordon | 071, 073 | 245.918, 245.293 |
| 1539 | Karl Gordon | 002, 027, 077 | 245.688, 245.757, 245.879 |
| 1512 | Néstor Espinoza | 001, 002, 003, 004, 005, 006, 007, 008, 009, 010, | 245.779, 245.806, 245.815, 245.832, 245.808, 245.801, 245.742, 245.656, 245.825, 245.828 |

In our previous report (Baines et al., 2023), we highlighted notable variations in the spectral trace's shape and position between visits, attributed to the GR700XD grism slightly over- or under-shooting its command position by a few fractions of a degree when inserted into the optical path. Furthermore, we illustrated how these variations could be modeled using simple linear models (see Fig 1 in Baines et al., 2023). Our findings suggest that the traces appear to rotate around a central point on the detector. This rotation causes a noticeable shift on the spectral features of the extracted spectra which suggests it also influences the wavelength solution from visit to visit.

The current wavelength solution for the GR700XD grism was derived for spectral orders 1, 2, and 3 during commissioning using observations of TWA 33 (a star of spectral class M5.5e), HAT-P-14 (a star of spectral class F5V), and BD+60-1753 (a star of spectral class A0V) using their strong and/or abundant absorption features in the near-infrared (Filippazzo et al., 2023). However, this model does not account for the pixel-wavelength shifts between observations mentioned above. Instead, offsets were estimated and applied to remedy a joint solution. In this Technical Report, we aim to study and characterize the changes in the wavelength solution. We conducted an experiment through a calibration program and included archival data from other calibration and commissioning programs.

## 3. Data Collection and Processing

For our wavelength calibration scheme, we focus on observations of the A-type star BD+60-1753, which has a significant number of repeat observations owing to several ongoing calibration activities. Additionally, we proposed and executed calibration Program 1512 (PI: Espinoza) to determine the GR700XD wavelength solution for orders 1 and 2 by obtaining 10 repeated observation of BD+60-1753. Each observation was performed subsequent to a Target Acquisition (TA) exposure, thereby necessitating the pupil wheel to transition between the CLEAR positions (used required for TA exposures in SOSSFAINT mode) and the GR700XD grism position. This strategy enabled us to sample 10 different pupil wheel positions for the NIRISS/SOSS mode since, as noted above, each time the wheel moves it causes the actual position of the pupil wheel to slightly deviate from the commanded position. We expand our dataset with observations of the A-type star from programs 1091 (PI: Martél; 1 observation), 1536 (PI: Gordon; 2 observations), and 1539 (PI: Gordon; 3 observations) for a total of 16 observations. These datsets were previously analyzed and published in Baines et al. (2023). The pupil wheel position angles—measured through the PWCPOS keyword—for each observation and program are presented in Table 1.

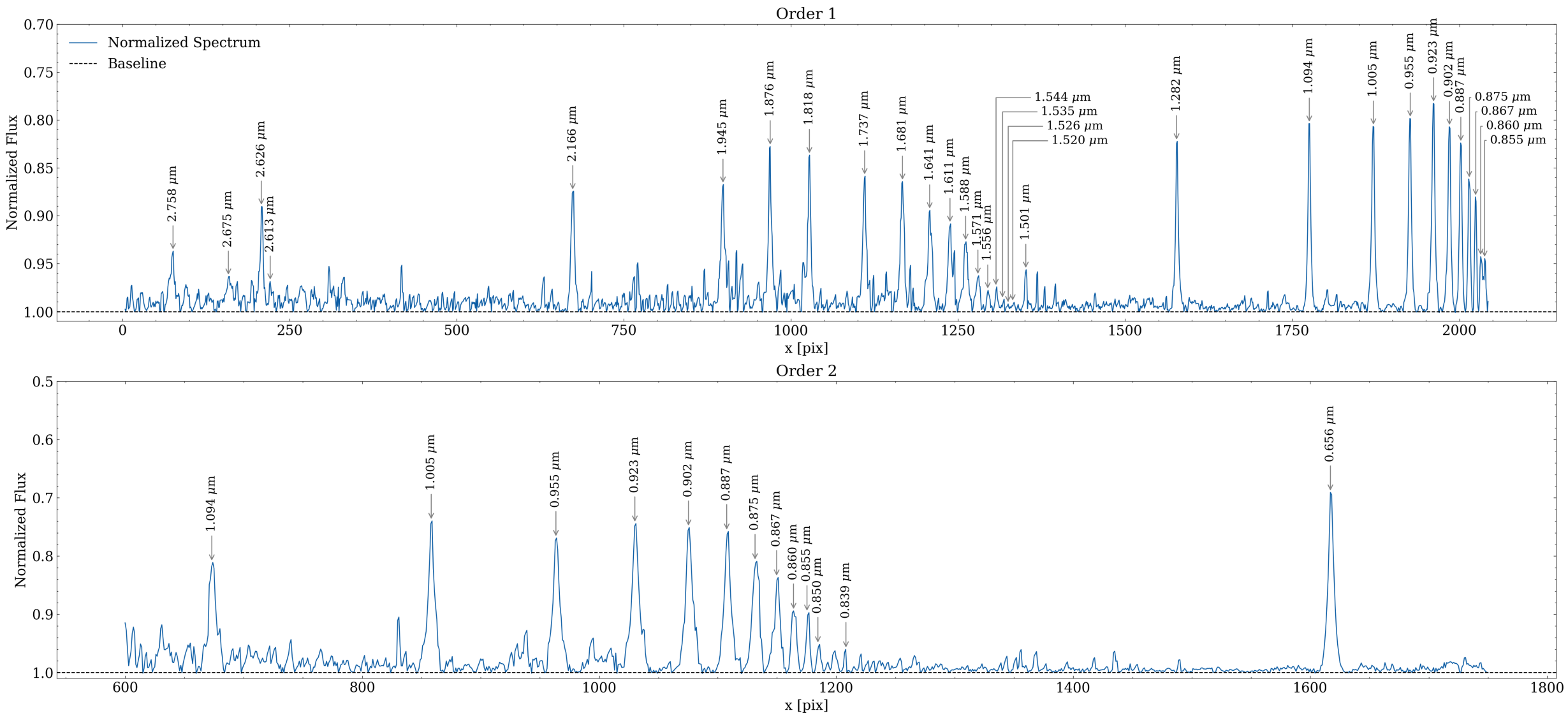

**Figure 1: Continuum-normalized spectra of BD+60-1753 for orders 1 and 2 with the y-axis inverted resulting in the absorption lines appearing like emission. Labels indicate prominent hydrogen absorption features. Refer to text for further details.**

We acquire the level-2a data products (i.e., the `_rateints.fits` files) for the Program 1512 SOSS observations from the MAST (Mikulski Archive for Space Telescopes) database. For each observation, we apply the same procedure outlined in Baines et al. (2023, see their Section 4) where we 1) derive the median integration image to produce a high signal-to-noise image and 2) measure the spectral trace positions for orders 1 and 2 in the resultant image. Also, we compile the median images along with the order 1 and 2 traces found previously in Baines et al. (2023) from programs 1091, 1536 and 1539 associated with the A-type star under study in this work. No traces from Order 1 were rejected. However, one of the order 2 traces from Program 1536 (observation number 071) was discarded due to contamination which resulted in significant deviations in the measured trace positions toward the long-wavelength end (i.e., pixel columns less than 1000) where a number of the spectral features are present.

We utilized the `getSimpleSpectrum` method from the *transitspectroscopy* Python package (Espinoza, 2022) for spectral extraction, employing a box aperture extraction routine with apertures of 20- and 15-pixels for orders 1 and 2, respectively. A continuum normalization was performed to enhance our capability of identifying and measuring spectral features. For spectrum normalization, the Baseline Asymmetric Least Squares (Baseline ALS) algorithm (Baek et al., 2015) is used to estimate and remove the continuum after applying a median filter to a spectrum. Figure 1 shows an example of one of the normalized spectra of BD+60-1753 for orders 1 and 2 using the Baseline ALS method, showcasing several Hydrogen absorption lines.

## 4. Wavelength Calibration

As noted in Baines et al. (2023), because of the slight visit-to-visit differences in pupil wheel positions for the GR700XD, the wavelength solution varies by a few pixels from visit to visit. We assume these PWCPOS-dependent wavelength solution shifts might be composed of two components: (1) a global shift of the spectrum, plus (2) individual/local wavelength-dependent shifts. In order to simplify our analysis and perform an initial exploration of the size of each of those shifts in our spectral datasets, we first aim to find and study global pixel offsets to each spectrum. We achieve this by employing the signal alignment algorithm developed by Pearson et al. (2019), which employs a cross-correlation strategy to align two signals: a reference and a target, within a specified region of interest. This algorithm has been successfully demonstrated in ground-based spectroscopy for systematic wavelength calibration. The algorithm offers both coarse and fine alignment strategies. For our purposes, we selected the coarse alignment to determine the global shift of each spectrum relative to a reference spectrum. Our reference spectrum was chosen based on its PWCPOS value, which is closest to the nominal position of 245.76 degrees (Martel et al. 2016, 2021, 2022), as shown in Figure 2.

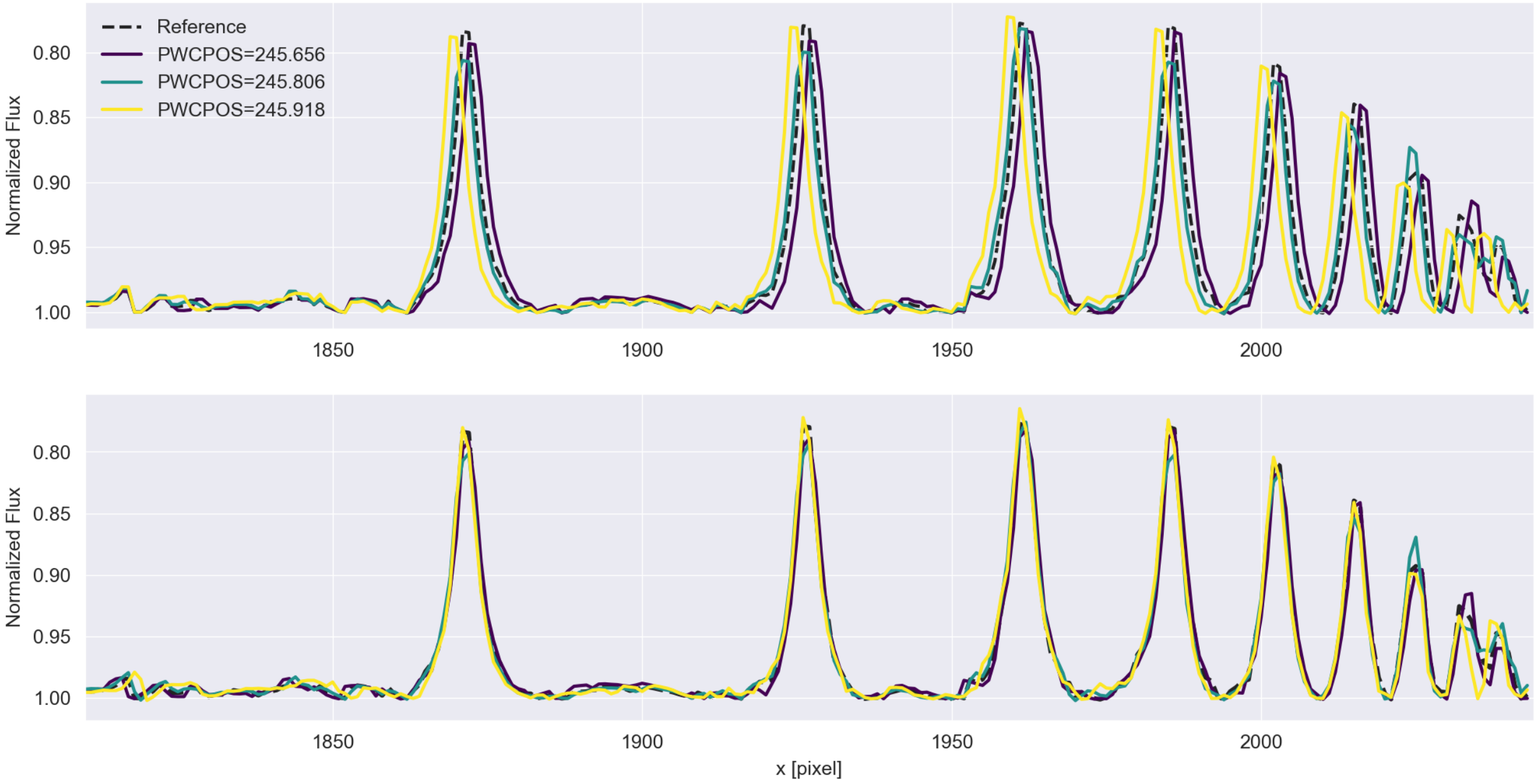

**Figure 2: Here we show the a the before (top) and after (bottom) signal alignment for spectra at three PWCPOS values as the colored lines. The black dashed line corresponds to the spectrum associated with the PWCPOS value closest to the commanded position. Prior to alignment, we observe how the GR700XD influences the dispersion solution, causing the lines to shift from right to left. The spectra line up nicely after applying spectral alignment algorithm.**

The measured global shifts of the order 1 spectra are shown in Figure 3, where the reference spectrum's PWCPOS is given by the red dot while the blue star denotes the commanded PWCPOS value. The global shifts of the spectra can vary as much as about 2.5 pixels from visit to visit: as much as 1 pixel in the positive direction for PWCPOS values lower than the 245.76 degrees reference value, and as much as 1.5 pixels in the negative direction for larger PWCPOS values. It is interesting to note that the relationship between this global pixel shift and PWCPOS is not linear. Given the small number of samples, we employ a Leave-One-Out cross validation analysis and find a degree 3 polynomial is favored over a linear model. The spectra for order 2 also showed a similar pattern. Further analysis and data may be required to verify this behavior.

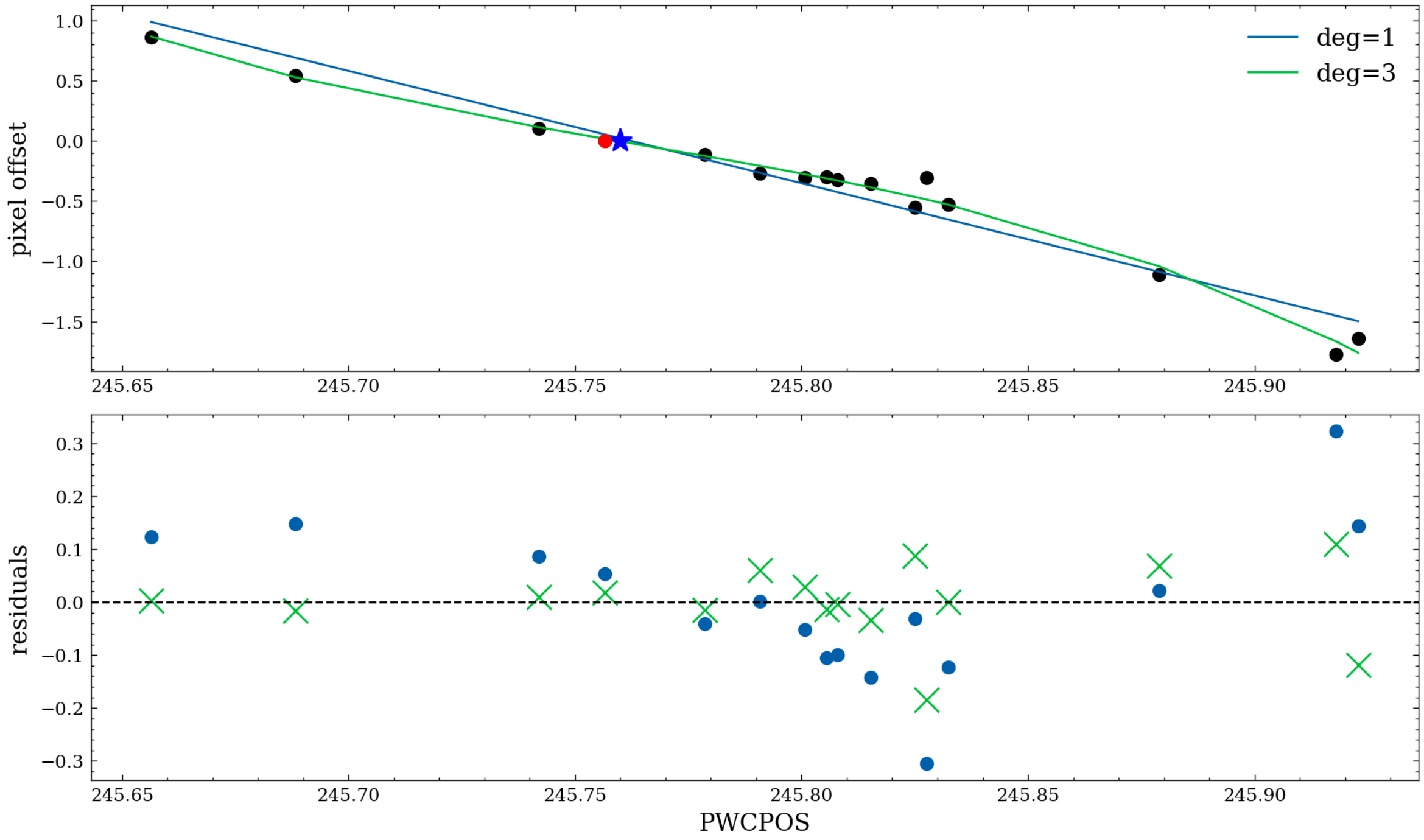

**Figure 3: This figure shows the resulting measured shifts for each of our sample of order 1 spectra for a given PWCPOS value. The red dot denotes the pupil position of the reference spectrum that all subsequent spectra are aligned to while the blue-star represents the nominal position angle which is assumed to be the zero-point. In addition, we show the fitted polynomial models for degrees 1 and 3.**

We estimated the line positions for each of the individual strong absorption features in each spectrum. We used the same list of Hydrogen lines at vacuum wavelength as in Filippazzo et al. (2023). They identified about 25 strong features for order 1 and 13 features for order 2 between 0.6 – 2.8 μm which served as our initial starting positions. We decided to include a few more hydrogen lines increasing the total number of features to 31 for order 1 where we added some weaker yet potentially useable spectral features. We use the previously determined line positions from Filippazzo et al. (2023) for orders 1 and 2 to predict the starting positions of the newly added line positions.

We subsequently improved our initial guesses by fitting a Voigt line profile to a narrow region spanning ±12 pixels around each line, centered on the predicted locations. This process was carried out using a bootstrap sampling scheme, enabling us to derive not only estimates of the line positions but also their associated uncertainties. As expected, the strongest absorption features were the ones with the smallest uncertainties, with weaker lines being more challenging to fit, which resulted in larger uncertainties. We found that lines near the red end of the spectrum (wavelengths > 2.0 μm) were particularly difficult to fit due to their low signal-to-noise ratio.

**Table 2: Summary of fitted polynomial regression models for spectral orders 1 and 2. We include the weighted and unweighted measured RMS error.**

| | Spectral Order | |
|---|---|---|
| | Order 1 | Order 2 |
| Scaler Transformation | Min/Max | Min/Max |
| Polynomial Degree | 5 | 3 |
| Fitted x range (pixel) | [73.63, 2039.74] | [672.14, 1619.42] |
| Fitted PWCPOS range (deg) | [245.656, 245.923] | [245.656, 245.918] |
| Fitted Wavelength Range (μm) | [0.854778, 2.758276] | [0.656470, 1.094116] |
| Number of $H_2$ Absorption Features | 31 | 13 |
| Unweighted RMSE (μm) | $(5.72 \pm 1.71) \times 10^{-4}$ | $(2.42 \pm 0.51) \times 10^{-4}$ |
| Unweighted RMSE (pixel) | $0.58 \pm 0.17$ | $0.52 \pm 0.11$ |
| Weighted RMSE (μm) | $(2.01 \pm 0.37) \times 10^{-4}$ | $(6.06 \pm 2.07) \times 10^{-5}$ |
| Weighted RMSE (pixel) | $0.21 \pm 0.04$ | $0.13 \pm 0.04$ |

We use the measured pixel positions of our Hydrogen lines to build a model that predicts the wavelength solution for each spectral order. To this end, we apply a polynomial regression scheme to train a model using the *scikit-learn* package (Pedregosa et al*., 2011)*. Our input features are the fitted x-pixel positions of the Hydrogen spectral features along with their respective PWCPOS offsets (the difference between the pupil wheel's actual position and commanded position). Our output/target feature is their associated assigned wavelength values. We applied a min/max scaler transformation to standardize our data, and split our data into an 80/20 train-test split prior to fitting. We use a linear regressor to fit the data and apply weights to each of our samples using inverse-variance weighting. We determine the best parameters for our model using a K-fold cross validation where we find that a polynomial of degree 5 and 3 for orders 1 and 2, respectively, best describe the data. Thus, our model for each order will take the form:

$$\lambda(x_1, x_2) = \sum_{i}^{N}\sum_{j}^{N} c_{i,j} x_1^i x_2^j$$

where $x_1$ and $x_2$ correspond the input features, N is the polynomial degree, and $c_{i,j}$ are the coefficients of the polynomial. We include all interaction terms in our model resulting a total of 20 coefficients for order 1 and 9 for order 2 with each having their own intercept terms.

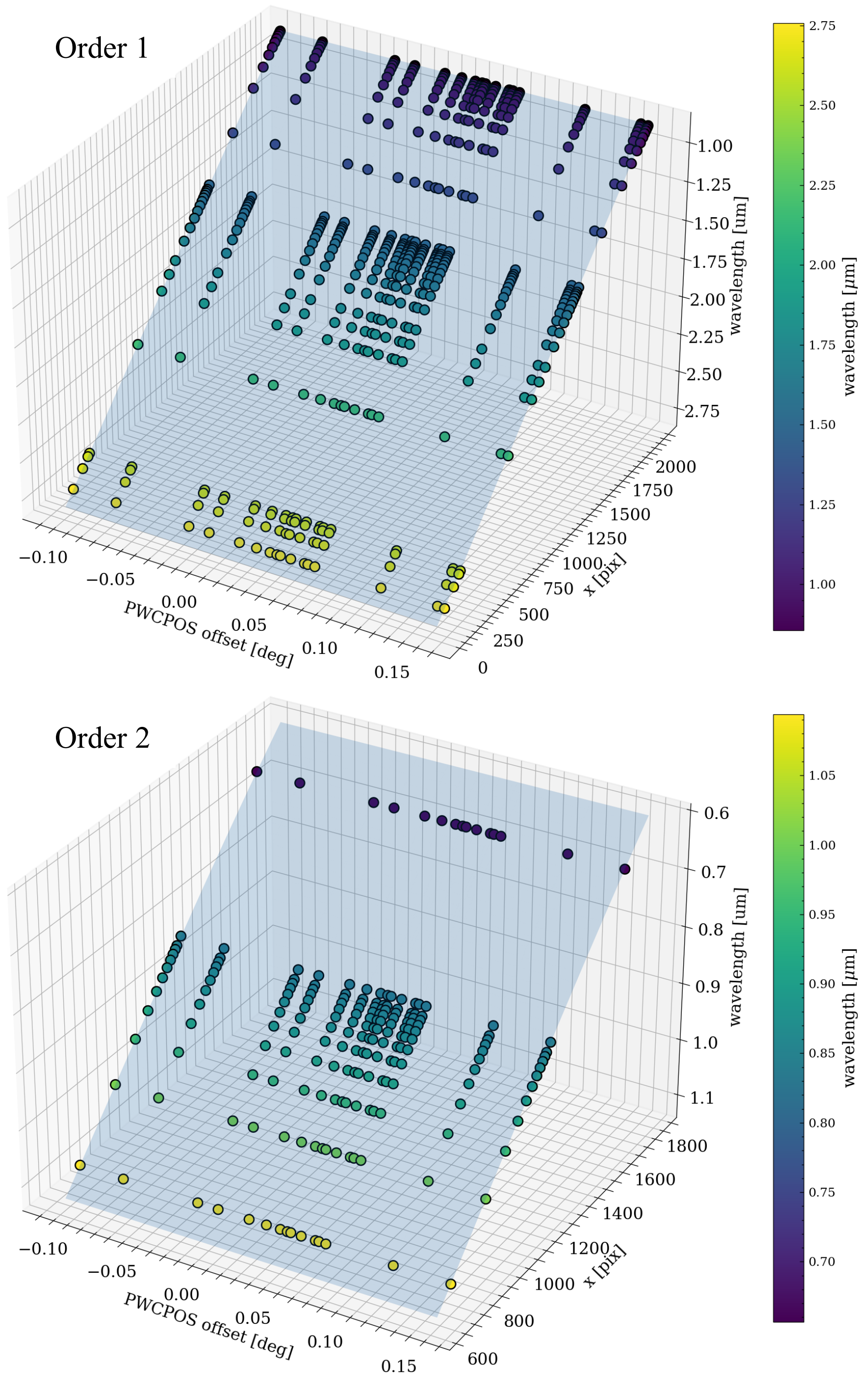

**Figure 4: This plot shows the polynomial regression model fit for orders 1 (top) and 2 (bottom) as the light blue plane. The scattered points show the respective line positions and their respective wavelength values observed at different PWCPOS offset values from the nominal position of 245.76 degrees.**

We fit our polynomial regression models (i.e., the calibrated wavelength solution) to our training datasets for both spectral orders 1 and 2, respectively; we show those results in Figure 4. In order to obtain a prediction error measure, we obtain the weighted mean square error (MSE) using our test dataset, and find an MSE of about 4.04 x $10^{-8}$ µm for order 1 and about 5.14 x $10^{-9}$ µm for order 2. The residuals (for the entire training and test dataset) of the fit for each model are shown in Figure 5. The average weighted-root mean square error (RMSE) of the entire (i.e., training and test) datasets suggests a sub-pixel accuracy with an RMSE of 0.21±0.04 pixels and 0.13±0.04 pixels for orders 1 and 2, respectively. Table 1 summarizes our wavelength calibration model and reports the weighted/unweighted RMSE accuracies for both spectral orders in wavelength- and pixel-space.

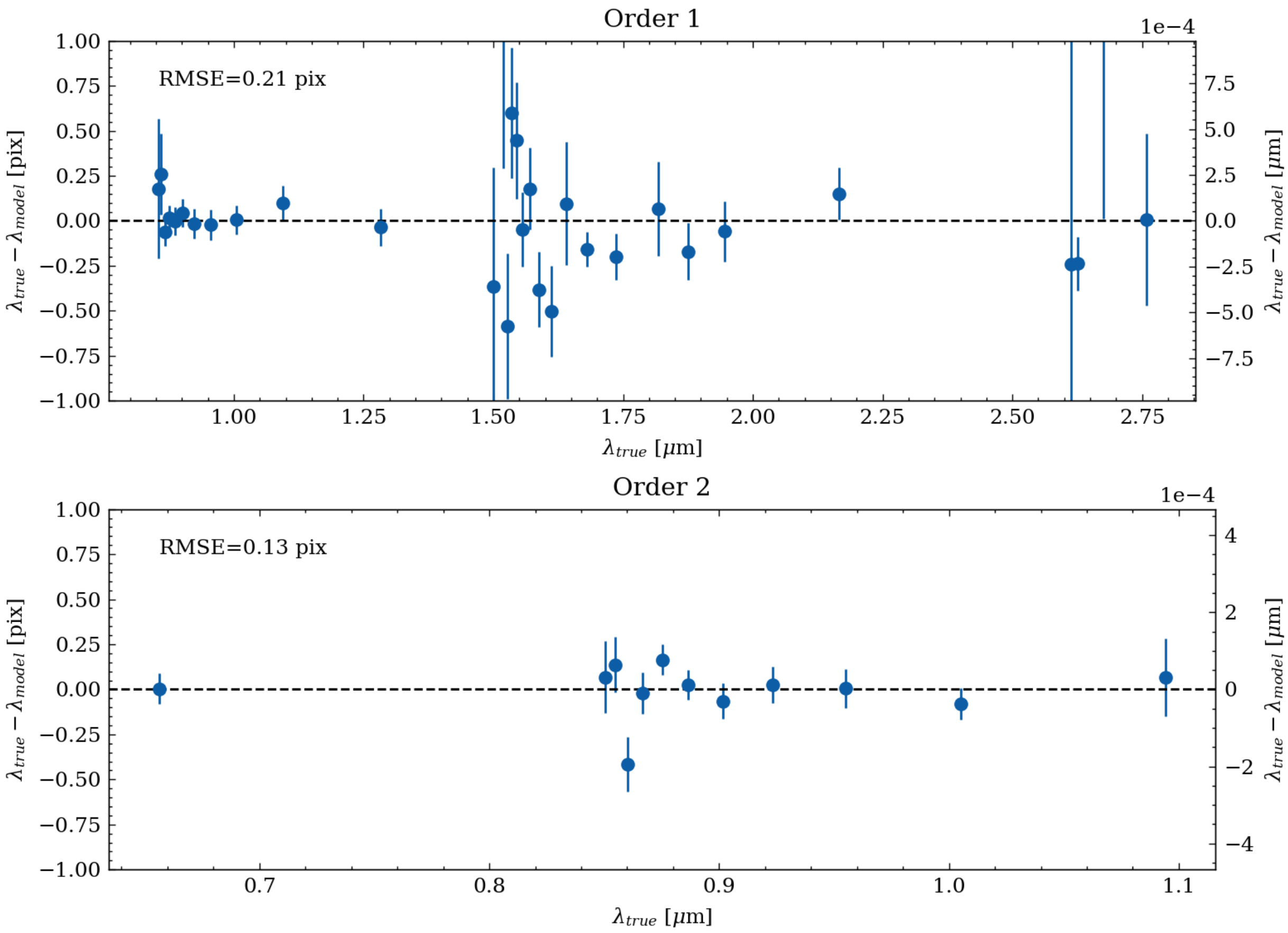

**Figure 5: Here we show the residuals of the trained polynomial regression model characterizing the wavelength solution for order 1 (top) and order 2 (bottom). Our model yields sub-pixel performance where we measure a weighted RMS of about 0.21 pixels and 0.13 pixels for orders 1 and 2, respectively.**

Having established the capability to predict the wavelength solution corresponding to a given PWCPOS value, we can proceed to determine the spectral resolving power for individual spectral orders. We calculate the spectral resolving power R on the basis of 2-pixel spectral resolution element given by the following,

$$R = \frac{\lambda}{\delta\lambda} = \left(2\frac{dln\lambda}{dx}\right)^{-1}.$$

The estimated spectral resolving power pertaining to a given wavelength solution in our sample is illustrated in Figure 6 alongside a CV3 reference model for comparison. Our model's calculated spectral resolving power exhibits good agreement with this reference model. Previous CV3 analyses determined that the dispersion for orders 1 and 2 are 0.96 +/- 2% (nm/pixel) and 0.467 +/- 2%, respectively (Albert et al 2023, SOSS III paper). Notably, our model yields similar measured dispersion values, approximating 0.98 nm/pixel at 1.84 μm for order 1 and 0.47 nm/pixel at 0.87 μm for order 2; again, in very good agreement with previous results.

Our model provides a robust means to predict the visit-to-visit wavelength solution in a given spectral order given the trace positions and PWCPOS offsets. Trace positions can be obtained for a given PWCPOS using the software introduced in Baines et al. (2013), PASTASOSS;

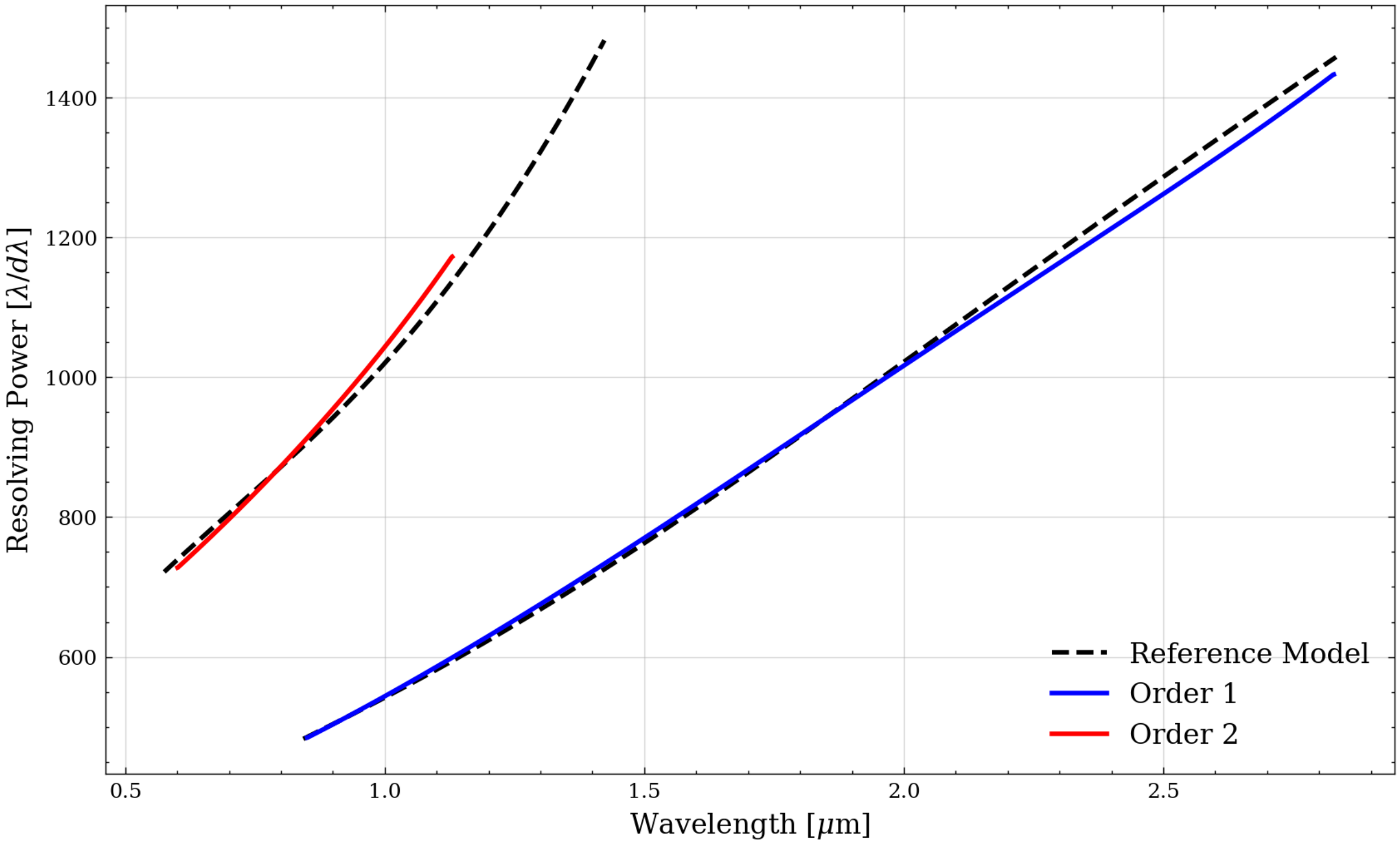

**Figure 6: Here we show the measured spectral resolutions derived from our polynomial regression models for spectral order 1 (blue) and 2 (red) in addition to a reference model based on measurements from CV3 ground-testing (NIRISS/GR700XD, JWST Documentation, 2016-2023).**

PWCPOS themselves can be obtained directly from the headers of JWST FITS products. In Figure 7, we illustrate the results of the visit-to-visit wavelength calibration for spectral orders 1 and 2 for NIRISS/SOSS observations which highlights the change in trace positions and wavelength shifts for a few reference wavelength values. As can be observed, while the effect is small, it is significant and particularly important for longer wavelengths in both Orders 1 and 2.

Our wavelength calibration model has been implemented in the PASTASOSS Python package, which has been developed to provide the community with a tool to quickly predict spectral order trace positions and their associated wavelength solution for a given PWCPOS value with the models described above and in Baines et al. (2013). We expect to improve and update those models in the future as more data is gathered, as well as to support trace and wavelength calibration of spectral Order 3.

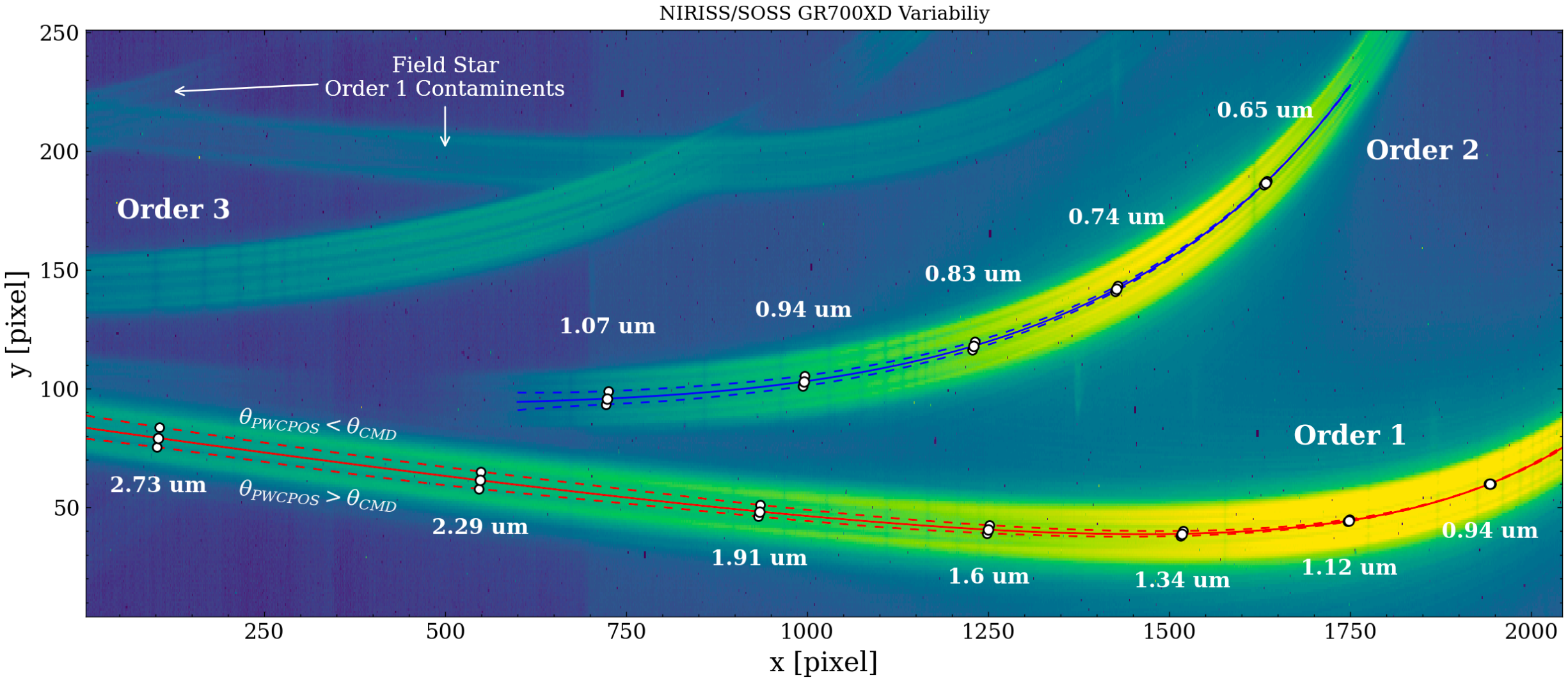

**Figure 7: Example NIRISS SOSS observation of BD+60-1753 (Program 1091), displaying the flux in log scale. The trace positions for spectral orders 1 and 2 are shown in red and blue lines, respectively, with solid lines representing the reference trace at PWCPOS=245.7909 degrees. The dashed lines represented the upper (below reference) and lower (above reference) limit in the bounded interval [245.6564, 245.9229] in degrees. The white dots represent a set of wavelengths of the reference as well as for the edge cases. The physical overlap of the 1st and 2nd order is seen on the left, corresponding to the red-end of both orders where trace positions deviate the most. In addition, we see the NIRISS SOSS background plus some field star contamination of 0th order and some 1st order spectra. The striped banding across the columns is from 1/f noise.**

## 5. Conclusion

We developed a model to improve the wavelength calibration of NIRISS/SOSS data for spectral orders 1 and 2 using data from a program (Program 1512, PI: Espinoza). This program was designed to monitor wavelength calibration variations with the pupil wheel position using calibration star BD+60-1753, which has strong Hydrogen absorption features. Data from other programs observing this star were used to supplement Program 1512 (PID 1091, PI: Martel; PIDs 1536, 1539, PI: Gordon). Our model accounts for visit-to-visit variations in the positions of SOSS spectra due to changes in the pupil wheel position of the GR700XD grism. We employed the use of polynomial regression to model and fit the pixel-wavelength relation of over 30 Hydrogen features to derive a model that yields sub-pixel performance for both spectral orders. We calculated the spectral resolving power for orders 1 and 2 and show that it is in line with analysis from ground testing. As more data become available, we plan to update our models as well as provide support for Order 3. Our wavelength calibration model can be accessed in the publicly available PASTASOSS Python package.

## 6. Acknowledgements

We thank the reviewers for their feedback on this report.